\documentclass[aps,pra,reprint,superaddress,showpacs]{revtex4-1}
\usepackage{hyperref}

\usepackage{soul,color}
\usepackage{amsmath,bm}
\usepackage{graphicx}
\usepackage[raggedright,hang,nooneline]{subfigure}
\usepackage{float}

\begin{document}


\title{Simulation of sympathetic cooling an optically levitated magnetic nanoparticle via coupling to a cold atomic gas}


\author{T. Seberson$^{1}$, Peng Ju$^{1}$, Jonghoon Ahn$^{3}$, Jaehoon Bang$^{3}$, Tongcang Li$^{1,2,3,4}$, F. Robicheaux$^{1,2}$,}
\affiliation{$^{1}$Department of Physics and Astronomy, Purdue University, West Lafayette, Indiana 47907, USA}
\affiliation{$^{2}$Purdue Quantum Science and Engineering Institute, Purdue University, West Lafayette, Indiana 47907, USA}
\affiliation{$^{3}$School of Electrical and Computer Engineering, Purdue University, West Lafayette, Indiana 47907, USA}
\affiliation{$^{4}$Birck Nanotechnology Center, Purdue University, West Lafayette, Indiana 47907, USA}

\date{\today}

\begin{abstract}
A proposal for cooling the translational motion of optically levitated magnetic nanoparticles is presented. The theoretical cooling scheme involves the sympathetic cooling of a ferromagnetic YIG nanosphere with a spin-polarized atomic gas. Particle-atom cloud coupling is mediated through the magnetic dipole-dipole interaction. When the particle and atom oscillations are small compared to their separation, the interaction potential becomes dominantly linear which allows the particle to exchange energy with the $N$ atoms. While the atoms are continuously Doppler cooled, energy is able to be removed from the nanoparticle's motion as it exchanges energy with the atoms. The rate at which energy is removed from the nanoparticle's motion was studied for three species of atoms (Dy, Cr, Rb) by simulating the full $N+1$ equations of motion and was found to depend on system parameters with scalings that are consistent with a simplified model. The nanoparticle's damping rate due to sympathetic cooling is competitive with and has the potential to exceed commonly employed cooling methods. 
\end{abstract}


\maketitle

\section{Introduction}
Cooling the motion of an optically levitated nanoparticle to the motional ground state has proven to be a formidable experimental challenge. Limiting the nanoparticle temperature is inefficient detection of scattered light, laser shot noise, and phase noise, among others. These limitations seen in conventional tweezer traps have sparked theorists and experimentalists alike to explore new and hybrid levitated systems that may offer alternative routes to the quantum regime. Passive/sympathetic cooling schemes involving coupling different degrees of freedom or nearby particles has been explored \cite{Liu:17,Stickler2016a,Arita2013,Arita:18,Ge2018,Ranjit2015}. Cavity cooling has had success \cite{Kiesel14180,PhysRevLett.122.123601,PhysRevA.100.013805} where strong coupling rates have been achieved through coherent scattering with the addition of a tweezer trap and allowed cooling to the lowest reported occupation number of $\overline{n} <1$ \cite{PhysRevLett.122.123602,Delic2020}. Even in its beginning stages, all electrical or electro-optical hybrid systems utilizing electronic circuitry are able to reach mK temperatures \cite{Goldwater_2019,PhysRevA.99.041802,doi:10.1063/1.5081045,PhysRevA.99.051401,PhysRevLett.122.223602} with one particular experiment reaching $\overline{n} = 4$ through cold damping \cite{PhysRevLett.124.013603}. The field has also recently seen magnetic particles and traps being investigated \cite{Hsu2016,PhysRevLett.119.167202,Zhange1501286,PhysRevLett.124.093602}, such as studying the dynamics of a ferromagnetic particle levitated above a superconductor \cite{PhysRevApplied.11.044041,Druge_2014,Hofer_2019}. 

In the spirit of promising new systems, this paper investigates a possible method of cooling the translational motion of an optically trapped ferromagnetic nanoparticle by coupling to a spin-polarized cold atomic gas. The coupling arises from the magnetic dipole-dipole interaction and allows significant energy exchange between the two systems. While the atom cloud is continuously Doppler cooled, energy is extracted from the nanoparticle through this energy exchange. The coupling of a nanoparticle to an atom cloud has been proposed previously with the coupling mediated by scattered light into a cavity \cite{Ranjit2015}. Sympathetic cooling a particular vibrational mode of a membrane was successfully demonstrated using a similar technique \cite{Christoph_2018,Joeckel2015,PhysRevLett.107.223001,PhysRevA.87.023816}, but with final temperatures well above the ground state. The scheme proposed here does not require optical cavities and has the potential to cool to the quantum regime. 

Simulations show that an atom cloud containing $10^6$ or more atoms is sufficient to achieve damping rates that are competitive with or exceeding that of cold damping or parametric feedback cooling. The theoretical cooling scheme proposed is best suited for, but not limited to, nanoparticle frequencies in the 100 kHz range or larger. 

This article is organized as follows. In Sec. \ref{approx_model} the theoretical proposal to couple a spin-polarized atomic gas to a ferromagnetic nanosphere through the magnetic dipole-dipole interaction is given. In Sec. \ref{simulation}, simulation results of the particle-atom cloud system with continuous atom Doppler cooling are provided with a discussion of the cooling results and extension to three dimensions. Lastly, a comparison with other cooling methods is discussed as well as experimental considerations.

\begin{figure}[h]  
  \centering
    \includegraphics[width=0.4\textwidth]{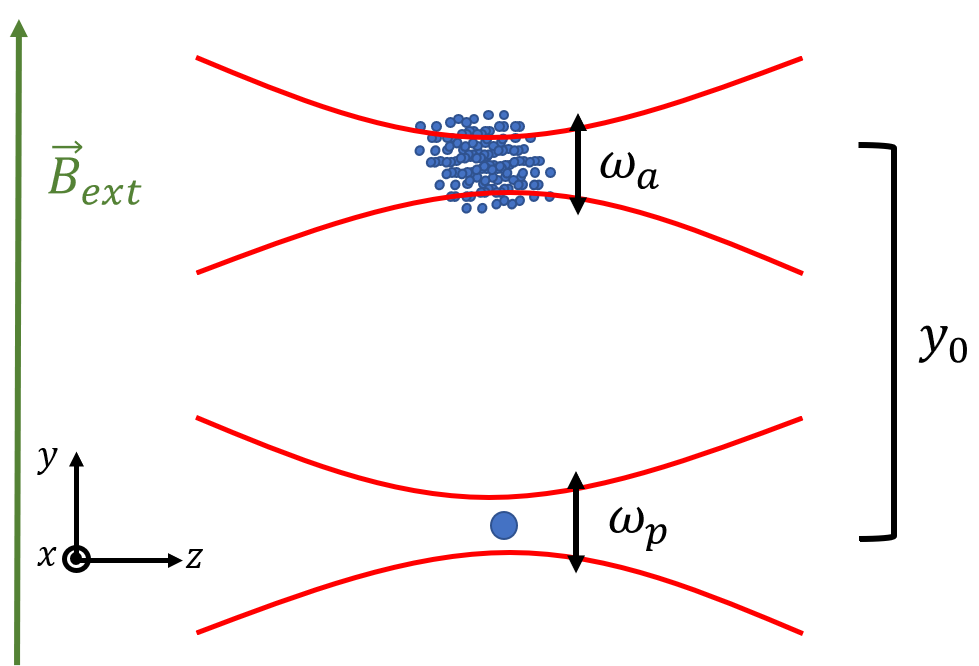} 
     \caption{Illustration of the proposed model. A ferromagnetic nanosphere is trapped at the focus of a Gaussian beam. The oscillation frequency for the nanoparticle in the $y$ direction is $\omega_p$. A cloud of $N$ atoms a distance $y_0$ away are trapped in a separate, far red-detuned dipole trap oscillating at frequency $\omega_a$ in the $y$ direction. An external magnetic field $\vec{B}_{ext}$ orients the magnetic moments of the nanoparticle and atoms. }  \label{fig:cold_gas_setup}
\end{figure}

\section{Model of the system \label{approx_model}}
The proposed physical system includes a ferromagnetic nanosphere of radius $R$ and mass $M_p$ harmonically trapped in the focus of a laser beam traveling in the $\vec{k}=\frac{2\pi}{\lambda}\hat{z}$ direction (see Fig. \ref{fig:cold_gas_setup}). A ferromagnetic sphere with dipole moment $\vec{m}$ produces a magnetic field \cite{jackson_classical_1999}
\begin{equation}
\vec{B}_{s}(r) = \left(\frac{\mu_0}{4\pi}\right)\left[ \frac{3\left( \vec{m}\cdot\hat{r}\right)\hat{r}}{r^3} - \frac{\vec{m}}{r^3} \right] , \label{eq10}
\end{equation}
where $\vec{r}$ is directed outwards from the center of the sphere. The sphere's moment will align along the $y$-axis if a constant, uniform magnetic field $\vec{B}_{\rm ext} = B_0\hat{y}$ is present, and a field distribution described by Eq. (\ref{eq10}) will surround the particle. 

A distance $y_0$ above the focus of the nanoparticle trap, a cloud of $N$, non-interacting, spin-polarized atoms each with dipole moment $\vec{\mu}_a=-\mu_a\hat{y}$ and mass $M_a$ are trapped in a far red-detuned dipole trap with oscillation frequency $\omega_a$. The total particle-atom cloud potential energy including the repulsive interaction $U_{\textrm{int},j} = -\vec{\mu}_{a,j}\cdot\vec{B}_s(r_j)$ for each atom $j$ is
\begin{align}
U = \frac{1}{2}M_p\omega_p^2r_p^2 + \sum^N_{j=1}\left( \frac{1}{2}M_a\omega_a^2r^2_{a,j} + U_{\textrm{int},j} \right), \label{eq11} 
\end{align}
where $r_{a,j}$ ($r_p$) is the atom (particle) position. If both the atoms and the nanoparticle undergo small oscillations compared to the distance separating them, $y_0\gg (r_p, r_{a,j})$, the interaction $U_{\textrm{int},j}$ is quasi-one-dimensional 
\begin{equation} 
U_{\textrm{int},j} \approx g/ \left(y_{a,j}+y_0-y_p \right)^3, \label{eq11.5}  
\end{equation}
where $g = 2\mu_a|\vec{m}|\mu_0/4\pi$ defines the interaction strength. The $N+1$ equations of motion for the $y$ degrees of freedom are
\begin{align}
\ddot{y}_p &= -\omega^2_p y_p - \sum^N_{j=1} \frac{3g/M_p}{ \left(y_{a,j}+y_0-y_p \right)^4 },  \label{eq51} \\ 
\begin{split} 
\ddot{y}_{a,1} &= -\omega^2_{a} y_{a,1} +  \frac{3g/M_a}{ \left(y_{a,1}+y_0-y_p \right)^4 }, \\ \label{eq52}
\qquad \qquad &... \\  
\ddot{y}_{a,N} &= -\omega^2_{a} y_{a,N} +  \frac{3g/M_a}{ \left(y_{a,N}+y_0-y_p \right)^4 }. \\ 
\end{split}        
\end{align}
The focus from here on will be the dynamics associated with the $y$ degree of freedom. The equations of motion for the $x$ and $z$ degrees of freedom have minimal coupling and are therefore largely harmonic oscillators. Extension to three dimensions is possible by placing the atom trap at a distance $\vec{r}_0=\langle x_0, y_0, z_0 \rangle$ while preserving anti-parallel atom and nanoparticle dipole orientations. For simplicity, the analysis in this paper focuses on one dimension. Further discussion of three dimensional cooling can be found in Sec. \ref{discussion}.

In what follows, the aim will be to study the possibility of removing motional energy from the nanoparticle sympathetically by continuously cooling each atom. Doppler cooling was the method of choice for cooling the atoms in this paper. Under Doppler cooling each atom experiences a momentum kick $\hbar k$ in the $\hat{k}$ direction if a photon is absorbed followed by a kick of the same magnitude in a random direction after spontaneous emission. The probability for absorption in each time interval $dt \ll 1/\omega_{a}$ is $\mathcal{P} = \mathcal{R}dt$ with absorption rate \cite{FootC.J.2011Ap}
\begin{equation}
\mathcal{R} =  \frac{\Gamma\Omega ^2/4}{\left( \Delta + \vec{v}_{a,j}\cdot\vec{k}\right)^2 + \Omega^2/2 +  \Gamma^2/4 } .\label{eq109}
\end{equation}
Here, $\Omega = \Gamma \sqrt{r/2}$ is the Rabi frequency, $\Gamma$ the decay rate, $r = I/I_{sat}$ the saturation intensity ratio, and $\Delta$ the laser detuning. The resulting atom velocity after this time interval can be written as
\begin{equation}
\dot{y}_{a,j}\left( t + dt  \right) = \dot{y}_{a,j} \left( t  \right) + \frac{\hbar k}{M_a} \begin{cases}
    \text{sgn}(\hat{k}) \pm 1,& \text{absorbed } \\
    0,              & \text{otherwise} \end{cases}  .\label{eq110}
\end{equation}

Other factors contributing to the dynamics of the trapped nanoparticle are collisions with the surrounding gas molecules and laser shot noise heating. For the vacuum chamber pressures used in atom trapping, $\sim 10^{-8}-10^{-9}$ Torr, the affects due to laser shot noise dominate that of the surrounding gas. After a time interval $dt$ the nanoparticle velocity becomes
\begin{equation}
\dot{y}_p(t+dt) = \dot{y}_p(t)+ \sqrt{2\dot{E}_Tdt/M_p}\mathcal{W}(0,1) ,\label{eq111}
\end{equation}
where $\dot{E}_T$ is the translational shot noise heating rate \cite{PhysRevA.102.033505} and $\mathcal{W}(0,1)$ is a random Gaussian number with zero mean and unit variance. 

Section \ref{simulation} presents the results of simulating Eqs. \eqref{eq51} to \eqref{eq111} with Eqs. \eqref{eq109} and \eqref{eq110} modeled using a Monte Carlo method. Subsections \ref{linear_couple} and \ref{cooling} explore the dynamics of Eqs. \eqref{eq51} and \eqref{eq52} analytically under a linear coupling approximation both with and without atom damping. 

\subsection{Dynamics under an approximate linear coupling \label{linear_couple}}
In the regime $y_0\gg (y_p, y_{a,j})$, Eq. (\ref{eq11.5}) may be expanded
\begin{equation} 
U_{\rm int,j} \approx \frac{g}{y^3_0} \left[  1 +  \frac{3}{y_0}\left(y_p-y_{a,j}\right) +  \frac{6}{y_0^2}\left(y_p-y_{a,j}\right)^2 + ...  \right].       \label{eq12}
\end{equation}
Keeping only terms to second order in Eq. (\ref{eq12}) and defining the center of mass of the atom cloud as $Y_a \equiv \frac{1}{N}\sum_{j=1}^N y_{a,j}$, the equations of motion may be written as
\begin{align} 
\ddot{y}_p &= -Na_p -\left( \omega^2_p+N\Omega^2_p\right)y_p + N\Omega^2_pY_a, \label{eq15} \\[1em]
\ddot{Y}_a &= \mkern15mu a_a - \left( \omega^2_a+\Omega^2_a\right)Y_a + \Omega^2_ay_p,      \label{eq16}
\end{align}
where $a_i= \left( 3g/M_iy_0^4\right)$, $i=(p,a)$, is a constant acceleration that shifts the equilibrium position of the oscillator and $\Omega^2_i = \left( 12g/M_iy_0^5\right)$ is a coupling constant as well as a frequency shift in the harmonic potential. The $\Omega_{i}$ here are different from the Rabi frequency defined in Eq. \eqref{eq109}.    

Provided the condition $y_0\gg (y_p,y_a)$ is maintained so that higher order terms in Eq. (\ref{eq12}) do not contribute, the nanoparticle will exchange energy with the atom cloud due to the linear coupling in Eqs. (\ref{eq15}) and (\ref{eq16}). Retaining predominantly the lower order terms is only possible for nanoparticle temperatures much smaller than room temperature as the atoms' positions will increase drastically as they exchange energy with the particle. The condition may also be satisfied if the motion of the atoms is continuously cooled via a cooling mechanism such as Doppler cooling, which will be discussed in the  next subsection. 

Due to the frequency shifts, $\Omega^2_i$, the nanoparticle and/or atom cloud trap frequencies need to be tuned to resonance for coherent energy exchange $\omega^2_a+\Omega^2_a = \omega^2_p+N\Omega^2_p =\omega^2_0$. While on resonance, the rate at which energy is exchanged from the particle to the atoms is solved for by finding the normal mode frequencies of Eqs. (\ref{eq15}) and (\ref{eq16}) and identifying the beat frequency. This "exchange frequency" is 
\begin{equation}
    f_{\rm exch} = \frac{\Omega_p\Omega_a\sqrt{N}}{\pi\omega_0}. \label{eq17}
\end{equation}
Note that while the overall force due to the magnetic dipole-dipole interaction was chosen to be repulsive, the dynamics are similar for an attractive interaction since the energy exchange effect is independent of the sign of the linear coupling term. Thus, although the atom cloud may be uniformly spin-polarized through the external magnetic field and optical pumping \cite{PhysRevLett.83.1311,GRIMM200095,Schmidt:03}, there is no loss of coupling if atoms undergo spin flips on time scales greater than the oscillation period. 

\subsection{Sympathetic cooling with linear coupling\label{cooling}}
If each atom in the cloud is continuously Doppler cooled, motional energy can be removed from the nanoparticle. For the calculations in this section, Doppler cooling is modeled using Langevin dynamics and is valid for the time scales of consideration here, $\Gamma\gg\omega_0$. 

Doppler cooled atoms experience an effective damping force $F_{D,j}=-\alpha \dot{y}_{a,j}$ with damping rate $\Gamma_a=\alpha/M_a = \hbar k^2 I/\left(I_0M_a\right) $ when tuned to reach the Doppler temperature $T_{\rm min}=\hbar\Gamma/2k_B$ \cite{FootC.J.2011Ap,Schmidt:03,Castin:89}. Excluding the constant accelerations in Eqs. (\ref{eq15}) and (\ref{eq16}), the equations of motion with Doppler cooling on the atoms as well as laser shot noise on the nanoparticle become
\begin{align} 
\ddot{y}_p &=  - \omega^2_0 y_p + N\Omega^2_p Y_a +\xi_{\rm SN}(t), \label{eq19} \\[1em]    
\ddot{Y}_a &= - \omega^2_0 Y_a + \Omega^2_a y_p -\Gamma_a \dot{Y}_a + \xi_{\rm DC}(t)/\sqrt{N},      \label{eq20}
\end{align}
where $\xi_{\rm SN}(t)$ accounts for fluctuations due to laser shot noise and $\frac{1}{N}\sum_{j=1}^N F_{a,j}(t)/M_a\rightarrow \xi_{\rm DC}(t)/\sqrt{N}$ are fluctuations due to spontaneous emission during Doppler cooling.

With the atoms continuously Doppler cooled, the final temperature of the nanoparticle, $T_{p}$, under sympathetic cooling can be estimated. One method is through integration of the power spectral density (PSD) since $T_{p}\propto \int S_{yy}(\omega)d\omega = \langle y^2_p \rangle$. Fourier transforming Eqs. (\ref{eq19}) and (\ref{eq20}) and solving for the nanoparticle's displacement in the frequency domain, $\widetilde{y}_p = \mathcal{F}\{ y_p \} $, gives
\begin{equation}
\widetilde{y}_p = \left[  \frac{\sqrt{N}\Omega^2_p \widetilde{\xi}_{\rm DC}(\omega)}{\Delta} +  \widetilde{\xi}_{\rm SN}(\omega)   \right] \left[   \frac{1}{ \delta^2(\omega) -N\Omega^2_p\Omega^2_a/\Delta}  \right], \label{eq21}
\end{equation}
where $\Delta =  \delta^2(\omega) + i\Gamma_a\omega $ and $\delta^2(\omega) = \omega^2_0 - \omega^2$. Shot noise adds an overall constant to the noise floor of the spectrum and near $\omega\sim\omega_0$ the affects are negligible, $\widetilde{\xi}_{\rm SN}(\omega) \ll N\Omega^2_p \widetilde{\xi}_{\rm DC}(\omega)/\Delta$, and may therefore be omitted. Using the single sided noise spectrum $|\widetilde{\xi}_{\rm DC}(\omega)|^2 \rightarrow 4\Gamma_a k_B T_{\rm min}/M_a$, the PSD of the nanoparticle's displacement is
\begin{equation}
S_{yy}(\omega) = \frac{ N\Omega^4_p (4\Gamma_a k_B T_{\rm min}/M_a) }{  \delta^4(\omega) \left[ \left( \delta^2(\omega) - N\Omega^2_p\Omega^2_a/\delta^2(\omega) \right)^2 +\left( \Gamma_a\omega \right)^2 \right ] }. \label{eq22}
\end{equation}
Equation (\ref{eq22}) is exact in the absence of laser shot noise on the nanoparticle and has been confirmed through simulation of Eqs. (\ref{eq19}) and (\ref{eq20}) with and without $\xi_{SN}(t)$. The PSD is well described by two peaks located at $\omega^2_{\pm} = \omega^2_0 \pm \sqrt{N}\Omega_p\Omega_a$ in the weak coupling regime $\omega^2_0>\sqrt{N}\Omega_p\Omega_a$. In the strong coupling regime, $\omega^2_0<\sqrt{N}\Omega_p\Omega_a$, the $\omega_{-}$ mode becomes unstable and the particle heats exponentially. Estimates of the final temperature of the nanoparticle through numerical integration of Eq. (\ref{eq22}) are given in Sec. \ref{Results}.

A second measure of the nanoparticle's approximate final temperature can be found by comparing the relative heating and cooling rates. In the weak coupling $\Gamma_a\gg \Omega_p\Omega_a/\omega_0$ and underdamped regimes $\omega^2_0 \gg \Gamma^2_a$, the average cooling power is 
\begin{equation}
\langle P \rangle \approx N k_B T_a \frac{\left(\pi f_{\rm exch}\right)^2}{\Gamma_a} , \label{eq50} 
\end{equation}
where $T_a=T_{\rm min}$ is the average temperature of an atom. From Eq. (\ref{eq50}) the cooling rate is identified as
\begin{equation}
\gamma_{\rm cool} = \frac{ 144g^2 N }{\omega^2_0 M_p \alpha y_0^{10} }.  \label{eq18} 
\end{equation}
Note that the particle cooling rate in Eq. \eqref{eq18} depends on the number of atoms $N$, the atom cooling rate $\alpha$, as well as the magnetic interaction strength $g = \left(2\mu_a|\vec{m}|\mu_0/4\pi\right)\propto M_p$. Equation \eqref{eq18} shows that slower atom cooling is beneficial for faster cooling of the nanoparticle. However, it is important that $\alpha$ remain large enough so that the atoms remain in the regime $y_0\gg y_{a,j}$ and do not escape the trap as a result of heating. 

Since the final temperature of the nanoparticle will be limited by the atoms' Doppler Temperature, $T_{\rm min}$, we may infer a rate equation of the form
\begin{equation}
\dot{T}_{p} \approx -\gamma_{\rm cool}\left( T_{p} - T_{\rm min} \right) + \dot{T}_{\rm SN}, \label{eq91}
\end{equation}
where $\dot{T}_{\rm SN}$ is the heating rate due to laser shot noise. Equation (\ref{eq91}) yields an equilibrium temperature of $T_{p} = T_{\rm min} + \dot{T}_{\rm SN}/\gamma_{\rm cool}$, numerical values of which are calculated in Sec. \ref{Results}.

\section{Simulations of the full system \label{simulation}}
\subsection{ System description}
%
 \begin{table*}[t]
 \centering
 \begin{ruledtabular}
 \begin{tabular}{c | c | c | c | c | c} %
 \textbf{Element} & 
 $\mathbf{M_{a}}$ (a.u.) & $\bm{\mu_{a}}/\bm{\mu_{B}}$  &  $\bm{\Gamma}/\bm{2\pi}$ (MHz)  &  $\bm{\lambda_{\rm line}}$ (nm) &  $\bm{T_{\rm min}}$ ($\mu$K) \\
  \hline
 Dy & 162.5 & 10 & 32.2 & 421 & 760\\
    \hline
 Cr & 52 & 6 & 5.02 & 425 & 124\\
  \hline
 Rb & 86.9 & 1 & 6.06 & 780 & 146\\
 \end{tabular}
 \end{ruledtabular}
  \caption{\label{table1} Relevant properties of the three species of atoms and their Doppler parameters. From left to right: the atom mass, magnetic moment, decay rate, Doppler line, and Doppler temperature.}
 \end{table*}
To determine the extent to which energy can be extracted from a nanoparticle through the scheme outlined in the previous section, several thousand simulations of the $N+1$ equations of motion were performed using the \textit{full, non-linear}, one dimensional $1/y^3$ potential in Eq. (\ref{eq11.5}) while continuously Doppler cooling each atom. Equations (\ref{eq51}) and (\ref{eq52}) were numerically solved using a fourth order Runge-Kutta algorithm. Laser shot noise kicks on the particle were implemented using Eq. \eqref{eq111} and Doppler cooling was modeled with a Monte Carlo method using Eqs. \eqref{eq109} and \eqref{eq110}. 

The nanosphere used for the simulations was composed of YIG with a radius $R=50$ nm, density $\rho=5110~\text{kg}/\text{m}^{3}$, index of refraction $n=2.21$ \cite{PhysRevB.9.2134}, and magnetic dipole moment $|\vec{m}|= \frac{1}{5}N_p\mu_B=4.05\times10^{-18}~\text{J}\text{T}^{-1}$ where $\mu_B$ is the Bohr magneton and $N_p$ is the number of atoms that make up the entire YIG nanoparticle. This expression for the magnetic moment is approximate, but is conservative compared to what is achievable experimentally \cite{MUSA20171135,doi:10.1063/1.4973199}. 

The simulated nanoparticle was trapped in a $\omega_p/2\pi=100$ kHz optical trap at the focus of a laser beam linearly polarized along the lab frame $x$-direction and propagating in the $z$-direction with a wavelength $\lambda=1550$ nm $\gg R$, power 150 mW, and focused by a $\text{NA}=0.6$ objective. The nanosphere was set with initial positions and velocities conforming to a Maxwell-Boltzmann distribution at a temperature of $T=1$ K. This temperature can be reached experimentally by first using parametric feedback cooling or cold damping before implementing sympathetic cooling \cite{Gieseler2012,Li2011}. Feedback cooling to a lower temperature before sympathetic cooling is necessary for reaching the desired pressures, to provide small amplitude oscillations, and to ensure that the atoms do not absorb too much energy that they escape their respective trap. The translational shot noise heating rate $\dot{T}_{\rm SN} = 2\dot{E}_T/k_B = 72.4$ mK/s was computed using the Rayleigh expression \cite{PhysRevA.102.033505}. Due to the frequency shifts $\Omega^2_i$ in Eqs. (\ref{eq15}) and (\ref{eq16}), the nanoparticle's frequency was shifted to match the frequency of the atoms while at the atom Doppler temperature. 

Three species of atoms, dysprosium, chromium, and rubidium, were used for separate simulations. Relevant properties for these atoms are listed in Table \ref{table1}. The atom cloud trap center was placed $y_0 = \lambda/3 = 516$ nm away from the nanoparticle trap's center. The atom dipole trap frequency $\omega_a/2\pi=100$ kHz remained unshifted. The atoms were initially set with velocities at their Doppler temperature and were continuously Doppler cooled. Doppler cooling parameters were set such that the atoms would reach their Doppler temperature $\Omega = \Gamma\sqrt{r/2}$, $r = I/I_{\rm sat}=0.1$, and $\Delta = -\Gamma/2$ (see Sec. \ref{approx_model}). Interactions between the atoms and atom loss from the trap were not included in the simulation. For the parameters used in this paper, the atoms (Dy, Cr, Rb) and particle are sufficiently in the weak coupling regime (see Eq. (\ref{eq22})) $\sqrt{N}\Omega_p\Omega_a/\omega^2_0 = (8.23, 6.07, 1.19)\times10^{-3}$ for $N=10^4$, respectively. 

From the parameters above, the beam waist of the particle trap is $\sim 800$ nm while the atom-particle separation distance is set at $516$ nm. While this distance is flexible, we envision the atom trap to be more tightly focused with a wavelength smaller than the wavelength used to trap the nanoparticle. Further, the particle oscillation amplitude is $\sim10$ nm at $T=1$ K, an order of magnitude smaller than the separation distance. The separation distance is foreseen to be the main experimental challenge as an increase in the separation by a factor of two decreases the nanoparticle damping rate by a factor of $2^{10}=1024$. 

\subsection{\label{Results}Simulation Results}
\begin{figure}[h] 
  \centering
    \includegraphics[width=.45\textwidth]{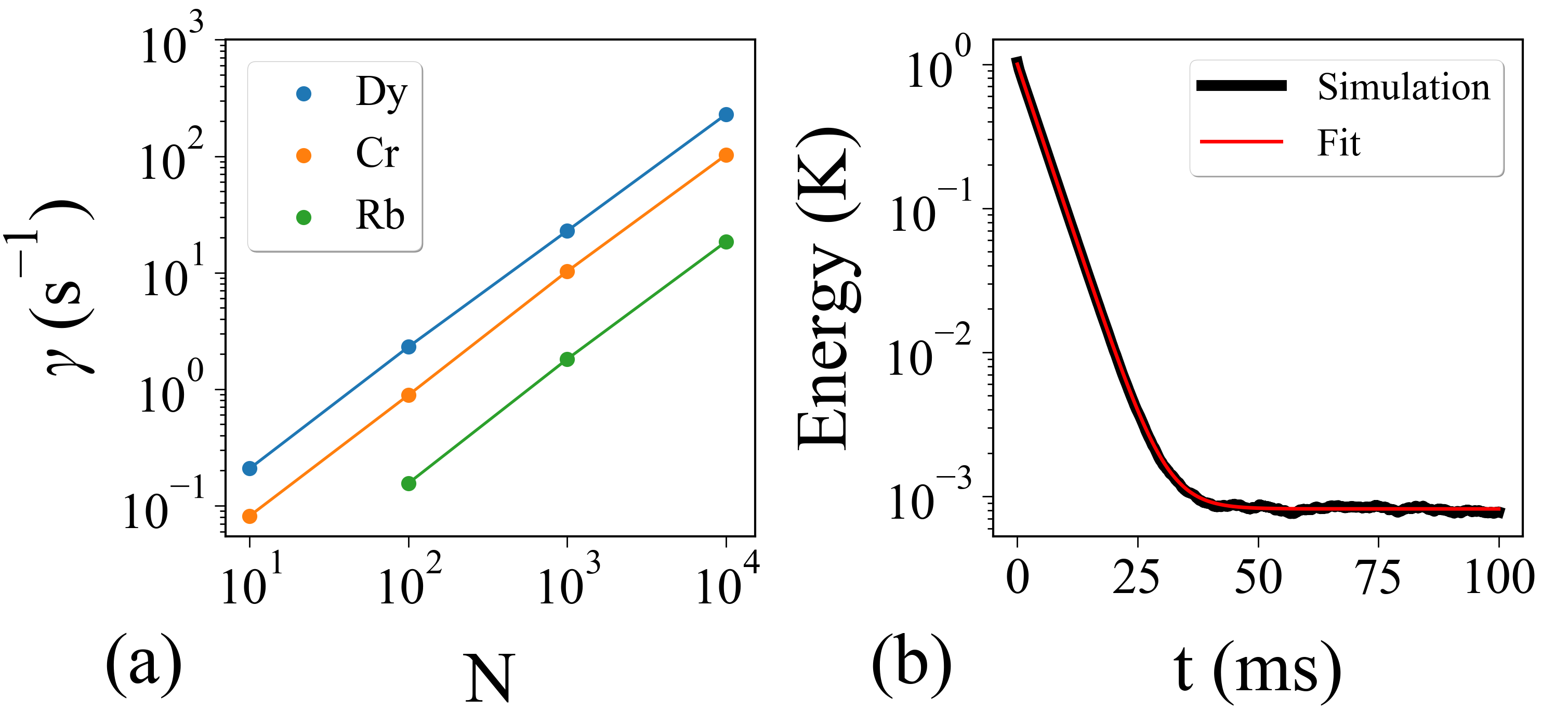} 
  \caption{(a) Nanoparticle cooling rate versus the number of atoms in the atom cloud. The rate is linearly proportional to the number of atoms and increases for species with larger magnetic moment $\mu_a$ as predicted by Eq. (\ref{eq18}). Only atom numbers that produced a statistically significant cooling rate were plotted. (b) Kinetic energy of the nanoparticle versus time for $N=10^4$ dysprosium atoms and fit to a decaying exponential. From the fit, $\gamma$ was extracted and used to plot (a).} \label{fig:rate_vs_n} 
\end{figure}
For each atom species, the energy removal rate of the nanoparticle, $\gamma$, was extracted for varying numbers of atoms in the trap, $N$, as seen in Fig. \ref{fig:rate_vs_n}. From several thousand averages, $\gamma$ was obtained by fitting energy versus time plots to a decaying exponential, $E(t) = A\exp(-\gamma t)+C$, for a given $N$ (see Fig. \ref{fig:rate_vs_n}(b)). It is to be noted that due to the long simulation times, most energy versus time plots did not allow for observation of the equilibrium temperature of the nanoparticle and were solely used for extracting $\gamma$. The final temperature of the nanoparticle can be greater than the atom Doppler Temperature as is the case in Fig. \ref{fig:rate_vs_n}(b) where the final average temperature is $794~\mu \text{K}$. This is attributed to shot noise heating as well as non-linear effects that contribute slight frequency mismatching between each atom and the nanoparticle. While the particle is resonant on the average, $\Omega_{a,j}$ for each atom $j$ depends on each atom's displacement amplitude and therefore changes slightly with time. The fact that the majority of atoms are out of phase with each other and the nanoparticle does not affect the results. 

From Fig. \ref{fig:rate_vs_n}(a), the cooling rate depends linearly on the number of atoms in the trap as Eq. (\ref{eq18}) predicts. As the particle exchanges energy with the atoms, each atom acquires a portion of the nanoparticle's energy. When $n$ more atoms are added to the trap, there are $n$ more chances for removing that energy through Doppler cooling. As Fig. \ref{fig:rate_vs_n}(a) shows no deviation from a linear dependence for larger $N$, the cooling rate may be extrapolated for larger $N$ values so long as $\omega^2_0>\sqrt{N}\Omega_p\Omega_a$. As the calculations were performed using the full $N+1$ equations of motion, simulation time was the only constraint from observing the effects for larger atom numbers, $N>10^4$. 

The average final temperatures reached for each atom species (Dy,Cr,Rb) was $794~\mu \text{K}, 406~\mu \text{K},$ and $158~\text{mK}$, respectively, for a simulation time of $t=100$ ms at $N=10^4$. Numerical integration of Eq. (\ref{eq22}) for the three atom species (Dy, Cr, Rb) at $N=10^4$ gives an approximated equilibrium temperature of $T_p \approx M_p\omega^2_0\langle y^2_p \rangle /k_B =  (2.002, 0.584, 0.572)~\text{mK}$, respectively. Since the particle is not a simple harmonic oscillator the values are approximate, but may serve as an upper bound for the particle temperature in experiments. The final temperature may additionally be estimated using the rate equation Eq. (\ref{eq91}), $T_{p} = (784, 256, 756)~ \mu$K at $N=10^4$ for (Dy, Cr, Rb), respectively, which shows better agreement with the simulation results for Dy and Cr. A simulation time of 100 ms was not long enough to observe the equilibrium temperature of the particle using Rb atoms. The expressions used to estimate the final temperature assume equilibrium has been reached, hence the discrepancy between the simulation temperature and the estimates. 

Besides the number of atoms in the trap, Eq. (\ref{eq18}) predicts that the cooling rate $\gamma_{\rm cool}\propto g^2 $ depends on the square of the magnetic coupling strength $g = 2\mu_a|\vec{m}|\mu_0/4\pi$. Using the data in Fig. \ref{fig:rate_vs_n}(a) and the values for $\mu_a$ in Table \ref{table1}, $\gamma \propto \mu^2_a$ is confirmed with a coefficient of determination $\textit{r}^2=0.997$. Together with the linear dependence of $\gamma$ on $N$, this confirms the ability to describe this non-linear system in an approximated linear coupling regime as was done in Sec. \ref{cooling}.

Note that $\gamma$ in Fig. \ref{fig:rate_vs_n} includes the cooling rate $\gamma_{\rm cool}$ as well as the competing shot noise heating rate $\dot{E}_T$. The two parameters that these two quantities share are the density, which is predominantly fixed, and the size (radius $R$) of the nanoparticle. Approximating $|\vec{m}| \propto R^3$ we find $\gamma_{\rm cool}\propto R^3$ while $\dot{E}_T \propto R^3$ shares the same $R$ dependence \cite{PhysRevA.102.033505}. However, shot noise heating is linear in time while the cooling is exponential, indicating that larger particles may provide faster cooling, but will not influence the final temperature of the nanoparticle. The influence of shot noise heating may be further reduced by cooling the degree of freedom in the laser polarization direction, $\hat{x}$, since the least amount of shot noise is delivered to that degree of freedom for a particle in the Rayleigh limit \cite{PhysRevA.102.033505}. The major source of heating in conventional levitated systems are collisions with the surrounding gas for pressures above $10^{-6}$ Torr with $\Gamma_{\rm gas}/2\pi\sim 10~\text{kHz}$. Cold atom experiments typically operate with $10^{-8}-10^{-9}$ Torr chamber pressures, making laser shot noise $\dot{E}_T/\hbar\omega_0 = 7.5$ kHz the dominant source of heating for the YIG particle in this proposal.  

\subsection{Discussion \label{discussion}}

The sympathetic cooling scheme can be extended to give three dimensional cooling by placing the center of the atom cloud a distance $\vec{r}_0 = \langle x_0, y_0, z_0 \rangle$ away from the nanoparticle and directing the magnetic field in that direction $\hat{B}_{\rm ext} = \hat{r}_0$ so that the dipole moments align. In this case all of the translational degrees of freedom would be coupled with one another, as well as with all of the translational degrees of freedom of the atoms. Despite the various couplings, simulations of the equations of motion up to the same order as in Sec. \ref{approx_model} show that this does not cause instability and allows cooling of each degree of freedom so long as different degrees of freedom are not resonant with one another, $\omega_{i,a}=\omega_{i,p}$, $\omega_{i,a}\neq \omega_{j,a}$ where $(i,j)=(x,y,z)$ and $i\neq j$. This condition is found experimentally since generally tweezer traps have different beam waists in the $x$ and $y$ directions causing the frequencies to be separated. 

For three dimensional cooling, it is possible to match all three frequencies. The radial frequencies ($\omega_x,\omega_y$) can be set by tuning the respective beam waists of the beam which can be controlled by either tightening the focus or tuning the lens. The axial frequency ($\omega_z$) is dependent on the wavelength and the $x,y$ beam waists non-linearly. The frequency of the atoms and the particle share the same dependence on these parameters. To match the atoms to the particle, each frequency has the same dependence on the trap intensity, scaling proportionally. 

To date, cold damping and cavity cooling by coherent scattering have proven to be the most effective methods of cooling a levitated nanoparticle with reported occupation numbers of $\overline{n} = 4$ and $\overline{n}<1$, respectively \cite{PhysRevLett.124.013603,Delic2020}. Similar to parametric feedback cooling, cold damping provides damping rates in the $\sim 1~\text{kHz}$ range while coherent scattering has yielded rates in the $10~\text{kHz}$ range \cite{PhysRevLett.124.013603,Ferialdi_2019,PhysRevLett.122.123602,Delic2020}. The results above indicate that atom numbers of the order $N\sim 10^5 (10^6) $, which corresponds to $ \gamma > 1$ kHz, would be sufficient for the nanoparticle to reach the atom's Doppler temperature for atom species with magnetic moment $\mu_a/\mu_B= 10 (1)$. The number of atoms that have been trapped experimentally is in the range $\sim 10^6-10^8$ for chromium, rubidium, and others \cite{Griesmaier2006,Ranjit2015,Schmidt:03}. The ground state energy of the nanoparticle is $T = \hbar\omega_0/k_B=5~\mu \text{K}$. Many of the commonly trapped atom species have unit magnetic moment, large $N$, and are able to reach the $\sim 1-10~\mu$K regime \cite{Lu2011,Gabardos2019}. Comparing with the energy removal rate found for Rb in Fig. \ref{fig:rate_vs_n}(a), these parameters are sufficient for motional ground state cooling of the nanoparticle. 

Admittedly, performing an experiment with the exact numbers as outlined in this paper may be a challenging task with currently technology. Optical dipole-traps allow for 100 kHz range trapping frequencies, large atom numbers, as well as low temperatures \cite{PhysRevLett.107.223001}. However, obtaining high enough densities to support an atom-nanoparticle separation of $<1 \mu\text{m}$ may be difficult on long time scales. To the best of our knowledge, atom densities of $10^{12}-10^{15}\,\rm{cm}^{-3}$ are achievable to date \cite{PhysRevA.102.013114,PhysRevA.101.033832,Newell:03,Ahmadi_2005,PhysRevA.71.021401}.  

The general idea of the proposed scheme is to sympathetically cool a nanoparticle utilizing the linear coupling found in the dipole-dipole interaction, but is not limited to the methods chosen here and allows for adaptability. For example, Doppler cooling as the atom cooling method was chosen for simulation simplicity while retaining physicality. Other atom cooling methods offer lower temperatures such as sideband cooling, Sisyphus cooling, or using a spin-polarized Bose-Einstein condensate, which are able to reach nK temperatures \cite{Lu2011a}. A charged YIG particle could also be trapped in an ion trap instead of an optical trap. 

Bose-Einstein condensates may have potential as sympathetic cooling candidates owing to their large densities ($10^{15}\,\rm{cm}^{-3}$) and low temperatures \cite{GRIMM200095}. Bose-Einstein condensates have been shown to reach submicron separations from a SiN surface \cite{PhysRevLett.104.143002,PhysRevLett.99.140403}. In the experiment of Ref. \cite{PhysRevLett.104.143002}, with $\omega_a=10$ kHz (5 kHz) and $N=10^3$, the radius of the Rb BEC was 290 nm (430 nm). The BEC was able to resonantly couple with a cantilever at separations of $\sim 1 \mu\text{m}$ with the cantilever at room temperature and no active cooling applied to the system. Further limiting the separation to the cantilever was the attractive BEC-surface interactive potential $\propto 1/r^4$ which distorted the trapping potential at small separations. The magnetic dipole-dipole interaction in this paper is $\propto 1/r^3$, has no concern with surface interactions, and assumes a system stabilized by active cooling. These considerations imply shorter separations are attainable. As BEC's are quantum in nature, a separate investigation of this possibility would be in order, though the methods would be similar to those outlined in this classical investigation.

\section{Conclusion\label{Conclusion}} 
A theoretical proposal to sympathetically cool a levitated ferromagnetic nanoparticle via coupling to a spin-polarized atomic gas was studied. While oscillating in their respective traps, the particle and atom cloud systems would be coupled through the non-linear magnetic dipole-dipole interaction. For sufficiently large separation between the particle and the atom cloud relative to their displacements, the nanoparticle and atom cloud would exchange energy with one another via the linear coupling term that is dominant in the magnetic force expansion. If the atoms are continuously Doppler cooled, energy would be able to be removed from the particle's motion. 

Simulations of the particle-atom cloud system were performed using the full, non-linear, magnetic dipole-dipole interaction for three species of atoms and varying numbers of atoms in the trap. The nanoparticle cooling rate was shown to be proportional to the number of atoms in the trap as well as the square of the magnetic moment of the atom, validating that it is possible to describe the dynamics using a linear approximation to the magnetic force. The rate at which energy is removed from the particle motion is significant for $10^4$ atoms in the trap when the atoms are continuously Doppler cooled. It is expected that the particle would reach the atom Doppler temperature as the number of atoms increases. This method of sympathetic cooling has potential to cool the nanoparticle to its motional ground state for atom species with lower Doppler temperatures. However, any atom cooling strategy that offers low enough temperatures should allow for motional ground state cooling if there are a sufficient number of atoms.

\begin{acknowledgments}

This work was supported by the Office of Naval Research (ONR) Basic Research Challenge (BRC) under Grant No. N00014-18-1-2371.

\end{acknowledgments}


\bibliographystyle{apsrev4-1}
\bibliography{Magnetic_coupling_to_atoms}

\end{document}